# CBC Approach for Evaluating Potential SaaS on the Cloud

Mrs. Dhanamma Jagli
Research Scholar, JNTU Hyd &
Assistant Professor, V.E.S.I.T,
University of Mumbai, India.
E-mail: dsjagli.vesit@gmail.com

Dr. (Mrs.) Sunita Mahajan,
Principal ICS-MET,
University of Mumbai, India.
E-mail: sunitam_ics@met.edu

Dr.N. Subhash Chandra,
Principal & Professor in
Dept. of CSE, HITS COE,
JNTU Hyderabad, India.
E-mail: principalhitscoe@gmail.com

**Abstract-**

**The cloud computing is evolving as a key computing platform for sharing resources like infrastructure, platform, software etc. This has proven to be an essential requirement for extending many existing applications. Software as a service (SaaS) is referred as on-demand software supplied by service providers in which software and associated data are hosted on the cloud and it can be accessed by service users using a thin client via a web browser. SaaS is commonly utilized and it provides many benefits to service users. To realize these benefits, it is essential to evaluate potential quality of SaaS, not only to the service users but also to the service providers. They have to evaluate their services against requirements of service users. The existing evaluation models are focusing only on quality attributes of SaaS. In this paper, a new evaluation model is proposed based on the data mining technique of Constraint Based Clustering (CBC). The proposed model gives emphasis on potential requirements of service users along with quality attributes of services.**

## I. INTRODUCTION

Cloud computing is a style of computing in which dynamically scalable and often virtualized resources are provided as a service over the Internet. One category of cloud service, SaaS is generally used and provides several benefits to service users. To recognize these benefits, it is crucial to evaluate the quality of SaaS and manage relatively higher level of its quality based on the evaluation result. In the precedent decade, the growth of web service technologies and the emergence of Service Oriented Architectures (SOAs) have added extremely to the mounting maturity of the Internet and the software industries. These advancements create it likely for software vendors to deliver effective software applications as web-based services using a new delivery model called Software-as-a-Service (SaaS) [2]. Software as a Service (SaaS) is a service delivery model that allows cloud service user to use the provider's applications running on a cloud infrastructure. In simple terms, SaaS is a model of software deployment everywhere an application is hosted as a service and provided to cloud service user across the Internet as per cloud service user requirements[2][3]. By means of eliminating necessitate installing and running the application on the customer side may be in the desktop computer, laptops, mobile device etc,

One of the categories of cloud services, SaaS is normally used and provides numerous benefits to cloud cloud service user [2] [9] [12]. To realize these benefits, it is essential to evaluate the software services on the cloud from customers prospective view and to manage relatively higher level of its quality based on the evaluation result, service providers should able to evaluate their quality of software as a service from customer prospective view and their constraints in while using software services. Therefore there is a highly demand for devising an evaluation model to evaluate SaaS quality as cloud service with customer requirements and constraints.

In present days, to understand the key metrics that are needed to optimize a SaaS business is the most important task. SaaS subscription businesses are more multifaceted than traditional businesses. In the SaaS world, there are a few key variables that make a huge difference to upcoming results. This paper is aimed at helping SaaS executives understand which variables actually stuff, as well as how to evaluate them and proceed on the results.

All existing SaaS evaluation models are based on quality attributes of software, but not focusing on actual customer requirements and their constraints, in fact customer retention is also an important factor. For example, small scale industry customer requirements for software may not be the same as individual service customer. In order to evaluate that SaaS evaluation model should also focus on the customer requirement status.

In addressing this problem, firstly in this paper, new evaluation model is proposed based on data mining CBC approach which considers customer requirements and constraints for evaluating quality of software and service for using SaaS. The remaining of the paper is organized as follows: section 2 discusses the related work, section 3 introduces the key features of SaaS and the general architecture of service provider, section 4 presents the CBC evaluation model. Finally, section 5 concludes this paper and discusses the future enhancements in our research work.

## II. RELATED WORK

**Evaluation :** It can be defined as "the procedure of determining the worth or significance of a development activity, policy or program to decide the relevance of objectives, the efficiency of design and execution, the effectiveness or else resource utilize, and the sustainability of outcome. An evaluation should facilitate the incorporation of lessons learned into the decision-making process of both provider and user.

**Existing Quality SaaS Evaluation models**
Jae Yoo Lee, Jung Woo Lee, Du Wan Cheun, and Soo Dong Kim work proposed a comprehensive model for evaluating quality of SaaS based on the key features of SaaS, derived quality attributes and metrics for the quality attributes. By using the proposed SaaS quality model, SaaS can be evaluated by service providers, in addition, the evaluation results are utilized as an indicator for SaaS quality management. The general architecture and considered quality attributes are as shown in the below Figure 1.

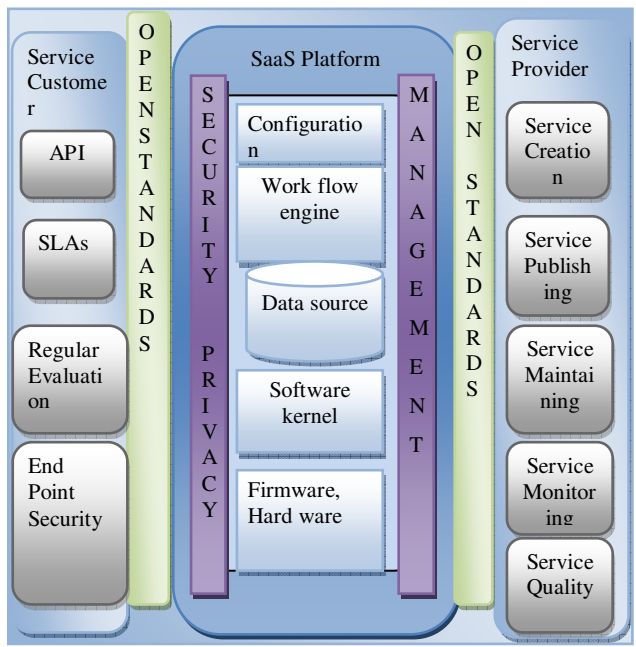

Figure 1: General Architecture for SaaS Service

Hence, it is required that this model is customized and extended to evaluate the potential quality in the software and service of SaaS.

Pang Xiong Wen and Li Dong work proposed new quality model which measure the safety measures, quality of service and software of the SaaS service, beginning from the perception of platform, service provider and service user separately. The quality evaluating model divides the SaaS service into four stages, including basic stage, standard stage, optimized stage and integrated stage. By using the quality model and evaluating model, service user can evaluate the provider service and the service provider can use it for quality management. Hence, it is required that this model is customized and extended to evaluate the potential quality in the software and service of SaaS.

Xian Chen, Abhishek Srivastava and Paul Sorenson work proposed a model framework for evaluating SaaS applications based on quality of service characteristics. Hence, it is required that this model be customized and extended to evaluate the potential quality in the software along with service.

Unfortunately, most current quality management approaches for SaaS services focus on the perspective of service providers, and thus do not fully take into consideration of customer requirements and incorporate the viewpoint of customers but often not in combination with the providers viewpoint. What is not present in the existing literature is an approach that adequately combines the perspectives of both provider and customer together with the nature of their ongoing business relationship with their constraints. According to the author's knowledge, working about how to evaluate SaaS software quality from service users prospective as for their potential requirement is not yet done.

## III. PROPOSED SAAS EVALUATION MODEL

### A. Potential software service evaluation.

A service system is a dynamic *configuration* of resources (people, technology, organizations, and shared information) that creates and delivers value between the provider and customer through service. In studying SaaS evaluation, it is focused on potential quality of software services and customer potential requirements and is motivated by two basic assumptions about the nature of service system delivery. Service system should consider customer requirements and their constraints.

❖ Service systems operate most effectively when both the service customer and the service provider understand and actively engage in the co-creation of value [2].

❖ Service system improvement is best achieved when the major software and service potential quality factors are mutually agreed upon, tracked, managed, analyzed and acted upon by the service customers and service providers.

- ❖ Service system can be modified as suitable for potential requirements of service users.

- ❖ Of resources (people, technology, organizations, and shared information) that create and deliver value between the provider and customer through service. In studying SaaS evaluation, it is focused on potential quality of software services and customer potential requirements and is motivated by two basic assumptions about the nature of service system delivery.

- ❖ Service system should consider customer requirements and their constraints.

- ❖ Service system improvement is best achieved when the major software and service potential quality factors are mutually agreed upon, tracked, managed, analyzed and acted upon by the service customer and service provider.

The objective of potential software and service quality management is to provide lower cost, better products and services and higher customer satisfaction. Traditionally, if a service provider understands what a customer wants from a service typically defined with a detailed specification based on the customer requirements and constraints, manages the variables in the service delivery process that can lead to deviation from specifications, and delivers the service in accordance with the customer's stated requirements, the service system is properly managing with respect to software and service potential quality [3].

In practice, however, a dynamic approach must be used in managing software and service potential quality due to continuous changes in the cost of service delivery, customer requirements and the emergence of new technologies. When existing customer expectations are not met, a new expectation benchmark must be set and service re-evaluation undertaken. The need is growing for evaluation models to assess software and service potential quality on an ongoing basis and to improve/accelerate decision-making related to the adoption of software services in general and SaaS applications in particular, given rapidly increasing adoption [6].SaaS and other recurring revenue businesses are different because the revenue for the service comes over an extended period of time (the customer lifetime). If a customer is happy with the service, they will stick around for a long time, and the profit that can be made from that customer will increase considerably. On the other hand, if a customer is unhappy, they will churn quickly, and the business will suffer.

### B. Basic features of SaaS

To define a quality model for evaluating SaaS, it is a prerequisite to identify key features of SaaS. From our rigorous evaluation on current references in CC [2] [7] [9], we identify key features as shown in below figure Fig 2.

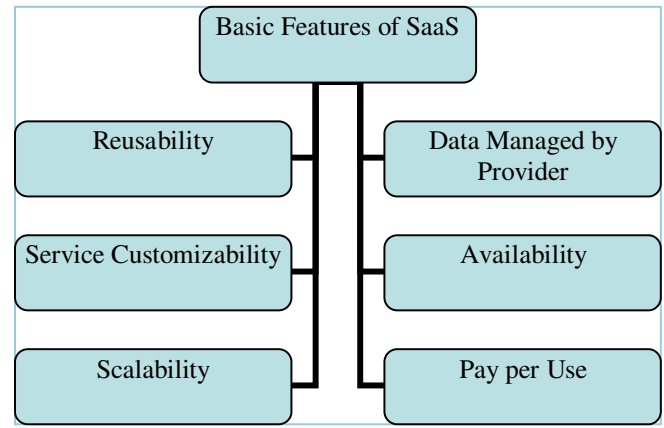

Figure 2: Basic features of SaaS

1) *Reusability:* is the ability of software elements to serve for the construction of many different applications. The fundamental underlying of cloud computing is to reuse various types of internet based services [2]. In case of SaaS, software itself is a target of reuse and it is delivered to service customer over the Internet. That is, one-to-many relationships are often used when delivering SaaS services.

2) *Data Managed by Provider:* SaaS is a model of software deployment whereby service providers license applications to cloud service user for use as a service on demand. Thus service providers are responsible for service installation and data management on their own server. Therefore, most data which service customers produce is stored on provider's data center and managed by provider. [10].

3) *Service Customizability:* Service customizability means the ability for services to be changed by service customers based on the individual requirements. This characteristic allows service providers to meet the different needs of each customer [11].

4) *Availability:* In cloud computing, the service consumers are able to access SaaS service from a Web browser via the Internet. Also, the customers do not have any ownership for the SaaS which is deployed and runs on the provider's server.

5) *Scalability*: In software engineering, scalability is a desirable property of a system, a network, or a process, which indicates its ability to either handle growing amounts of work or to be readily enlarged. Due to the black-box nature of cloud computing services, service customers cannot control resources which are utilized

6) *Pay per Use*: The expenses for SaaS are estimated not about purchasing an ownership of services, but about using services such as the number of service invocations or duration which services are utilized.

### C. Potential attributes to Evaluate SaaS

Back in the desktop software era, magazines ran software reviews in which the side-by-side comparisons of features took up an entire page. Buyers used these reviews to *shortlist* vendors, trying to anticipate which features they'd need over the next five years. Typically, the software with the most features won. Feature-it is ruled. No more. With software as a service, the focus has become whether the tool is good enough on day one and how well it will adapt over time. Indeed, in order to evaluate SaaS, the following seven potential attributes are critical as shown below in the Figure 3.

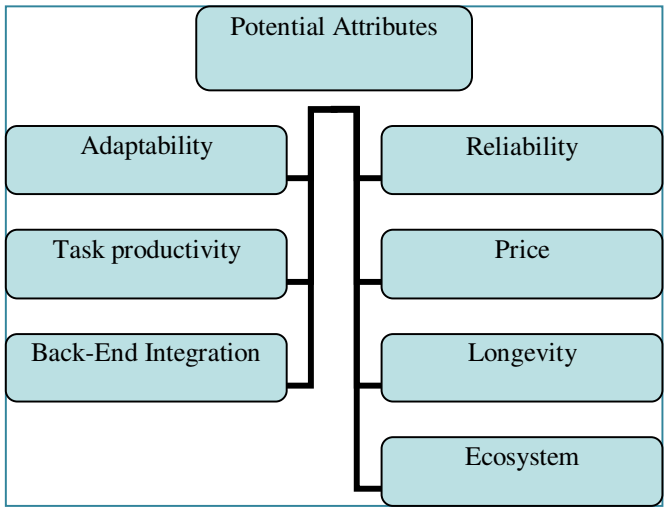

Figure 3: Potential Attributes to Evaluate SaaS

1) *Adaptability:* How easily application can be modify? This can be as simple as adding fields or building dashboards, or as advanced as a programming platform.
2) *Reliability:* How much system is dependable? This boils down to four things: Performance, availability, scalability and security.
3) *Task productivity:* How effectively users can accomplish their goals? How many cases-per-minute or entries-per-day can workers do, and how many errors do they make?
4) *Price:* How much will it cost — really? Because SaaS offerings are so varied in pricing, it's hard to compare them. A better model is to create several benchmark subscribers and compare upfront and ongoing costs for them.
5) *Back-end integration:* Can you plug it in to other things? Any enterprise SaaS offering will have to work with other systems, for everything from authentication to data sharing.
6) *Longevity:* How long will the SaaS Company be around, and what is an exit strategy? When a SaaS provider closes down, what about SaaS?
7) *Ecosystem:* How many third-party developers and integrators surround a particular platform with plug-ins

and add-ons, and how active are they? A vibrant ecosystem means a more extensible, flexible solution.

The key point, however, is that features on day one don't matter as much as the efficiencies and cost savings and squeeze out of the SaaS tool within 30 days of adoption — and how confident you are that those efficiencies and cost savings will endure .the most of current works are not for SaaS but for certain targets such as a conventional software or SOA based system. Due to the situation, it is hard to evaluate quality of SaaS and judge which SaaS is good. Therefore, our work provides a potential quality model to evaluate SaaS.

### D. ISO/IEC 9126 Quality Attributes

1) *Functionality* – The capability of the software product to provide functions which meet stated and implied needs when software is used under specified conditions.
2) *Reliability*- The capability of the software product to maintain a specified level of performance when used under specified conditions.
3) *Usability* – The capability of the software product to be understood learned, used and attractive to the user, when used under specified conditions.
4) *Efficiency*- The capability of the software product to provide appropriate performance, relative to the amount of resources used, under stated conditions.
5) *Maintainability* – The capability of the software product to be modified. Modifications may include corrections, improvements or adoptions of the software to changes in environment, and in requirements and functional specifications.
6) *Portability* – The capability of the software product to be transferred from one environment to another.

Based on ISO/IEC 9126, Quality attributes are applicable to conventional software but cloud software quality can not purely depend on these quality attributes because the nature of cloud software is not exactly same as convention software. So that mapping of quality attribute with SaaS features is not enough to evaluate potential services.

### IV. CONSTRAINT BASED CLUSTER APPROACH

#### A. Constraint Based Clustering

Constraint Based Clustering (CBC) is used to finds clusters that satisfy user-specified preferences or constraints find clusters that satisfy user-specified preferences or constraints. A constraint can express a user's expectation or describe properties of the desired clustering results, and provides an effective means for communicating with the clustering process. Constraint-based methods are used in spatial clustering for clustering with obstacle objects (e.g, considering obstacles such as rivers and highways when planning the placement of automated banking machines) and user-constrained cluster analysis (e.g, considering specific constraints regarding customer groups when determining the best location for a new service station, such as "must serve up to 200 high-value customers".

### B. Advantages of Constraint based clustering Approach

- It gives importance to user feedback: Users know their applications the best.
- It deals with less *parameters*, but more user-desired constraints.
- It deals with obstacle desired clusters.
- It is desirable to have user-guided (*i.e*, constrained) cluster analysis.
- It is used to find clusters that satisfy user-specified constraints — is highly desirable in many applications.

### C. Types of Constraints

Users often have a clear view of the application requirements, which they would ideally like to use to guide the clustering process and influence the clustering results. Thus, in many applications, it is desirable to have the clustering process take user preferences and constraints into consideration. Few categories of constraints are:

- Constraints on individual objects.
- Constraints on distance or similarity functions for example, Weighted functions, obstacles (*e.g*, rivers, lakes).
- Constraints on the selection of clustering parameters. For example No. of clusters.
- User-specified constraints for example contain at least 500 valued customers and 5000 ordinary ones.

### D. User-Specified Constraints for CBC

The cloud user requirements and constraints specified by him are playing a vital role in the process of SaaS evaluation on the cloud from service providers as well as service users. On the cloud computing environment service users can specify user constraints for using that particular software service. Few identified user constraints are as follows:

1. Number of instances that process the work in parallel.
2. Upper bound on Number of instances.
3. Total amount of work in the user's job.
4. Workload per instance.
5. Task length, time to process Workload per instance on a specific instance.
6. Budget per instance.
7. Users desired confidence in meeting budget.
8. Deadline on the user's job.
9. Desired confidence in meeting job's deadline.
10. User's bid on a Spot Instance type.

In this paper, constraint based clustering method is proposed to evaluate software services on the cloud this approach considers user specified constraints. Initially, it constructs clusters based on *k*-means clustering. Then, based on user constraints each cluster refines into micro cluster. Then, follows to identify deadlock; user specified constraints. The workflow of proposed CBC approach has shown below Figure 4.

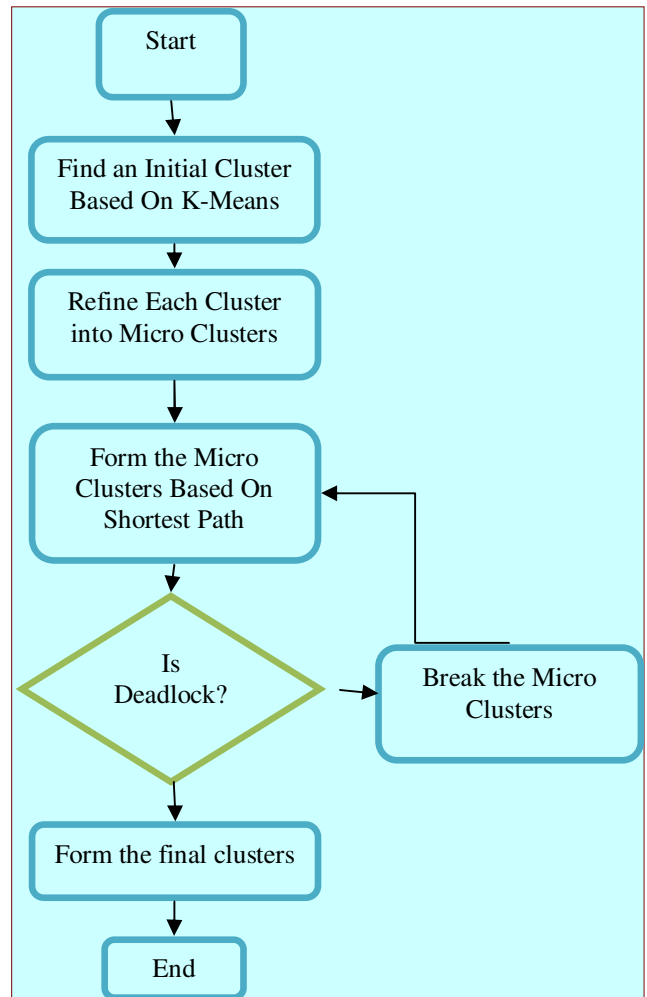

Figure 4: CBC Work Flow Stages

Constraint Based Clustering has been proven to be extremely useful. It has been applied not only to pattern discovery settings (e.g, for sequential pattern mining) but also, recently, on classification and clustering task. It appears as a key technology for an inductive database perspective on knowledge discovery in databases (KDD), and constraint based mining is indeed an answer to important data mining issues e.g, for supporting a prior relevancy and subjective interestingness, but to achieve computational feasibility. However, less research has done in this particular area. The sample data shown below in the table 1 is considered to analyze the result. The basic features of SaaS and its user rating for each attribute is taken from low to high on the scale factor 1 to 10, user constraints are considered as follows:

1. 50 instances that process work in parallel.
2. Max 100 number of instances required
3. 180 hours per month of work in the user's job.
4. Minimum requirement 48 hours per instance.
5. Maximum 5,000/ Rs- per instance can spend
6. Medium budget organization.
7. Only 30 days to be used 7 day trial period is required.

Table 1: Sample Data

| Tid | ReUsability | Customizability | Scalability | Availability | Data ManaGement | Pay perUse | Constraints |
|---|---|---|---|---|---|---|---|
| T100 | 2 | 2 | 4 | 2 | 3 | 5 | 6 |
| T101 | 3 | 5 | 3 | 3 | 4 | 4 | 7 |
| T102 | 4 | 4 | 2 | 4 | 5 | 8 | 3 |
| T103 | 5 | 5 | 5 | 4 | 5 | 2 | 9 |
| T104 | 2 | 3 | 4 | 5 | 4 | 5 | 5 |
| T105 | 3 | 2 | 3 | 5 | 3 | 3 | 6 |
| T106 | 4 | 4 | 2 | 4 | 2 | 4 | 7 |
| T107 | 5 | 5 | 2 | 4 | 1 | 5 | 6 |
| T108 | 5 | 5 | 3 | 3 | 1 | 5 | 5 |
| T109 | 4 | 3 | 4 | 3 | 3 | 4 | 4 |

The sample result of k-means clusters without considering user specified constraints and CBC clusters by considering user specified constraints are shown below, in the Figure 5.

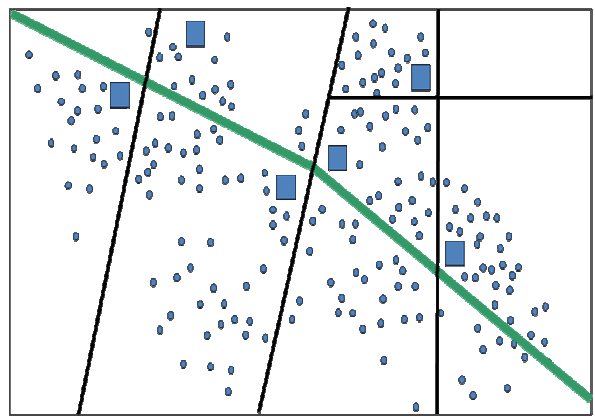

Figure 5: Sample CBC Results

### V. CONCLUSION AND FURTHER ENHANCEMENT

On the cloud computing SaaS is a one type of cloud service is emerging as a valuable reuse standard. It provides benefits to service customers without initial cost to purchase software, free of maintenance/updates, high availability, accessibility through Internet and pay-per-use pricing. As more and more SaaS service emerges, the quality of SaaS services are vastly different, so evaluating the quality of SaaS service becomes more essential issues both to customer and service provider.

In this paper, new approach for SaaS evaluation is proposed based on data mining constraint based clustering. This approach is very much useful to service providers and service customers to consider cloud user specified constraints. With some identified user constraint the approach is successfully analyzed. This proposed model would be very helpful to service providers to consider user specified constraints so that service customer retention can be achieved. It is also useful to service customers to evaluate SaaS quality with their potential requirements and constrains. In the future initially it was intended to complete the evaluation criteria for quality, potentiality with user constraints by applying, it could be automate data mining tools or by developing software tools to measure and evaluate SaaS service. Secondly, measuring cluster's quality is also aimed.